\documentclass[preprint,3p,times,twocolumn]{elsarticle}
\usepackage[utf8]{inputenc}
\usepackage[english]{babel}
\usepackage[hidelinks]{hyperref}
\usepackage{listings}
\usepackage{inconsolata}
\usepackage{color}
\usepackage{amsfonts}
\usepackage{amsmath}
\usepackage{braket}

\makeatletter
\def\ps@pprintTitle{%
 \let\@oddhead\@empty
 \let\@evenhead\@empty
 \def\@oddfoot{\@empty}%
 \let\@evenfoot\@oddfoot}
\makeatother

\begin{document}

\begin{frontmatter}

\title{Classical and Quantum Data Interaction in Programming Languages: \\ A Runtime Architecture}       

\author{Evandro Chagas Ribeiro da Rosa}
\ead{evandro.crr@posgrad.ufsc.br}
\author{Rafael de Santiago}
\ead{r.santiago@ufsc.br}

\address{Departamento de Informática e Estatística - INE PPGCC \\
         Universidade Federal de Santa Catarina - UFSC \\
         Florianópolis, Santa Catarina, Brazil}

\begin{abstract}
    We propose a runtime architecture that can be used in the development of a
    quantum programming language and its programming environment. The proposed
    runtime architecture enables dynamic interaction between classical and
    quantum data following the restriction that a quantum computer is available
    in the cloud as a batch computer, with no interaction with the classical
    computer during its execution. It is done by leaving the quantum code
    generation for the runtime and introducing the concept of futures for
    quantum measurements. When implemented in a quantum programming language,
    those strategies aim to facilitate the development of quantum applications,
    especially for beginning programmers and students.  Being suitable for the
    current Noisy Intermediate-Scale Quantum (NISQ) Computers, the runtime
    architecture is also appropriate for simulation and future Fault-Tolerance
    Quantum Computers.  
\end{abstract}



\end{frontmatter}

\section{Introduction}

A quantum computer uses quantum mechanics phenomena, such as superposition and
entanglement, to solve some problems faster than classical computers. Such a
claim was first proposed by Feynman \cite{Feynman1982} motivated by the
difficulty of simulating the evolution of quantum states. Later, confirmed by
the pioneer works of Shor \cite{Shor1997} and Grover \cite{Grover1997}. And,
recently demonstrated experimentally by Google \cite{Arute2019}. 

A quantum programming language describes instructions for a quantum computer.
It is a relatively new paradigm, but there are several implementations
\cite{Cross2017,Svore2018,Fu2019,Killoran2019,Khammassi2018,Heckey2015,
Javadiabhari2015,Smith2016,Steiger2018,Green2013,Wecker2014,Lapets2013} from
both industry and academy. The level of abstraction and quantum architecture
targets vary from each language. For instance, Q\# \cite{Svore2018} is a
high-level domain-specific language used to program quantum computers based on
quantum logic gates, and Blackbird \cite{Killoran2019} is a low-level language
used for continuous-variable quantum computation on photonic quantum computers. 

Today, we are in the Noisy Intermediate-Scale Quantum (NISQ) era
\cite{Preskill2018} marked by quantum computers with a few dozen of noise
qubits, where the commercial ones are accessible through the cloud,
\textit{e.g.}, the Amazon Braket service \cite{amazon}. Due to the low-fidelity
of the small number of qubits and the implemented gates, every quantum
execution needs to be as much optimized as possible, forbidding the execution
of a generic quantum code.  For example, to activate the best performance, the
quantum code that sums an arbitrary integer $x$ into a quantum superposition
should be specialized for every possible value of $x$.

In opposition to the NISQ computers, there are Fault-tolerant Quantum Computers
\cite{Devitt2013} with enough qubits to run expensive quantum error correction
(QEC) codes \cite{Devitt2013} on top of quantum applications. Nevertheless, we
are still far from them. Although even there, we may have the same restrictions
of accessibility (in the cloud, not locally) and the need to run highly
optimize quantum applications due to the high cost of the QEC codes. 

In this context, we propose a runtime architecture that can be used to both
develop quantum programming languages, or embed quantum programming in general
propose programming languages. The runtime architecture enables dynamic
interaction between classical and quantum data stored in qubits despite the
limitations that a quantum computer executes in bach, without interaction
between classical and quantum computers during the quantum execution, and that
the generated quantum code needs to be highly optimized for a specific
execution. More precisely, this dynamic interaction allows the loop of
classical information controlling quantum operations and quantum measurement
operation with classical information, all regardless of whether that data is
stored on the classic or quantum computer. 

With the proposed runtime architecture, a quantum programming language can be
developed as a general-propose programming language that supports the quantum
programming paradigm, and not a domain-specific programming language as must
quantum programming languages
\cite{Svore2018,Lapets2013,Wecker2014,Steiger2018}. This abstraction permits
the development of an application that takes advantage of quantum and classical
computers with a single and consistent programming language,  facilitating the
development of quantum applications, especially for beginning programmers and
students.

The proposed architecture has several components, where the main one is a
shared library that manages the quantum code generation, optimization, and
execution. Those components form a general framework for the construction of a
quantum programming environment with a simulator and optimizing compilers. 

The remainder of this paper is organized as follows. A brief introduction to
quantum computation is given in Section \ref{sec:computing}. An overview of
quantum programming and the constraints imposed on the proposed runtime
architecture are shown in Section \ref{sec:programming}.  The runtime
architecture and its component are detailed in Section \ref{fig:runtime}.  The
shared library features that enable the dynamic interaction between classical
and quantum data are presented in Section \ref{sec:lib}. And,  in Section
\ref{sec:conclusion}, there are our conclusions and final remarks. 

\section{Quantum computation}
\label{sec:computing}

This section introduces the concepts of quantum bit
(qubit), quantum measurement, quantum entanglement, quantum circuit, and
decoherence.

\begin{figure}[h]
    \centering
    \includegraphics[width=0.8\linewidth]{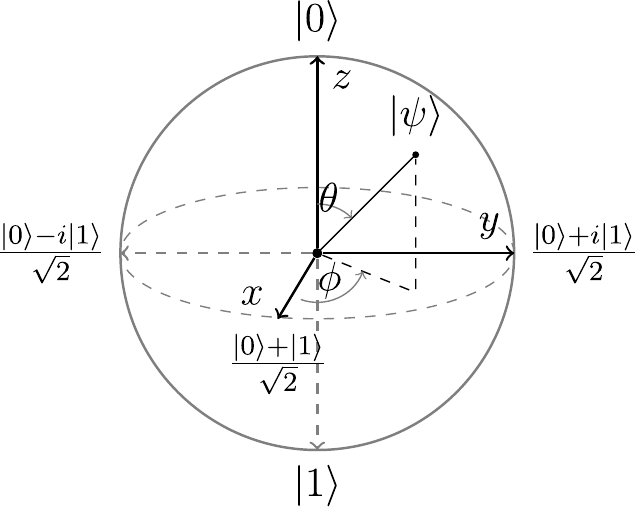}
    \caption{Bloch sphere used to visualize a single qubit.}
    \label{fig:bola}
\end{figure}

The basic unity of information used in quantum computation is the qubit. A
qubit (or quantum bit), alike to a bit, can be at states 0 or 1, denoted by
$\ket0$ and $\ket1$, but different from a bit, it can also be at both states at
the same time, in what we call a superposition. In a superposition, a qubit is
denoted by a convex sum,  \textit{e.g.}, $\ket\psi = \alpha\ket0+\beta\ket1$
where $\alpha, \beta \in \mathbb{C}$ and $|\alpha|^2+|\beta|^2 = 1$. It can be
visualized geometrically in the Bloch sphere, \textit{e.g.}, the qubit
$\ket\psi = \cos\frac{\theta}{2}\ket0 + e^{i\phi}\sin\frac{\theta}{2}\ket1$ of
Figure \ref{fig:bola}.

To extract information from a superposition, we need to perform a measurement
that will randomly return 0 or 1 according to the probability amplitude of the
qubit state, \textit{e.g.}, $\ket\psi = \alpha\ket0+\beta\ket1$ has the
probability $|\alpha|^2$ of return 0 and $|\beta|^2$ of return 1. When a
measurement is done, the qubit's superposition is destroyed, collapsing its
state. For example, if the measurement of $\ket\psi$ returns 0 or 1, it
collapses, respectively, to state $\ket0$ or $\ket1$.

Quantum entanglement is another important phenomenon for quantum computation.
When a set of qubits is entangled, we cannot fully describe every qubit
separately. It means that any operation on a single qubit will change the state
of all qubits in the set. For example, if we measure one of two maximally
entangled qubits, we know what will be the measurement result of the other
qubit.

The evolution of a quantum state is described by a unitary operation.
Consequently, any quantum operation, except measurement, is time-reversible,
and we cannot copy the state of a qubit \cite{Wootters1982}.  

\begin{figure}[h]
    \centering
    \includegraphics[width=.8\linewidth]{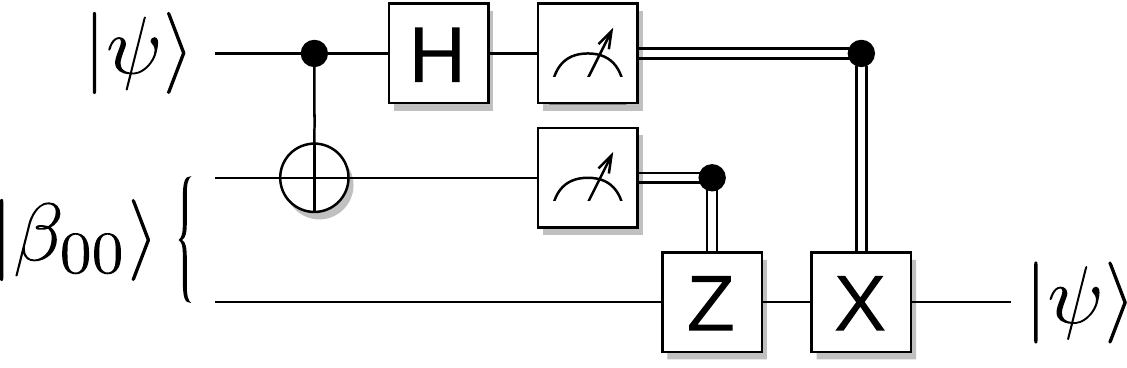}
    \caption{Quantum teleportation circuit, where $\ket{\beta_{00}} =
    \frac{\ket{00}+\ket{11}}{\sqrt{2}}$. The last two quantum
    gates Z and X are controlled by the measurement result of the top two
    qubits.}
    \label{fig:tele}
\end{figure}

A quantum circuit \cite{Nielsen2012} is a diagram that defines the evolution of
a quantum state through time. An example of quantum circuit can be seen in
Figure \ref{fig:tele}. A quantum circuit is executed from left to right by the
application of quantum gates and measurements. The therm quantum operation used
in this paper refers to any process (not necessarily a quantum circuit) that
changes the quantum state of a qubit.

Decoherence is the process of loss of information that a quantum computer
suffers due to its interaction with the environment, or the inaccurate
execution of quantum operations. It is the primary barrier for quantum
computation, limiting the time that a quantum computer can maintain the quantum
information, and consequently, the circuit deep (number of quantum gates). To
overcome this limitation, we can encode the quantum state in a quantum error
correction (QEC) code \cite{Devitt2013}. However, this approach adds a costly
overhead in the number of qubits and quantum gates used.

\section{Quantum programming}
\label{sec:programming}

In this section, a general overview of quantum programming languages is given
along with the quantum programming restriction and characteristics imposed on
the proposed runtime architecture.

Although it is acknowledged that an abstraction beyond the quantum circuit
would benefit the development of new quantum applications \cite{Svore2006}, the
abstraction of most quantum programming languages is equivalent to quantum
circuit.  It is particularly true for a set of quantum programming languages
\cite{Smith2016, Fu2019, Khammassi2018, Cross2017, Killoran2019} called quantum
instruction set or quantum assembly language. Those quantum programming
languages aim to be an intermediary representation for higher-level programming
languages and other quantum programming software \cite{qiskit, Smith2016,
Killoran2019}.  

Implement a quantum programming language as an embedded programming language is
a usual strategy \cite{Green2013, Steiger2018, Killoran2019, Wecker2014,
Javadiabhari2015} that can reduce development cost, and provide vast libraries
for the quantum programming language. However, the embedded programming
language will also be limited by the addictions of the base general-propose
programming language.

The most evident target for a quantum programming language is a quantum
computer. Although, most quantum programming languages provide constructions
that go beyond the NISQ computer processing power \cite{Svore2006, Green2013}.
This way, several targets are also provided. For example, classical simulators,
usually with less than 30 qubits; resource estimators to calculate the cost of
a particular quantum execution, usually, in numbers of qubits (circuit width)
and gates (circuit deep); and, quantum circuit drawers, use to generate a
graphic representation of a quantum execution, which is usually a quantum
circuit.

\begin{figure}[h]
    \centering
    \includegraphics[width=.8\linewidth]{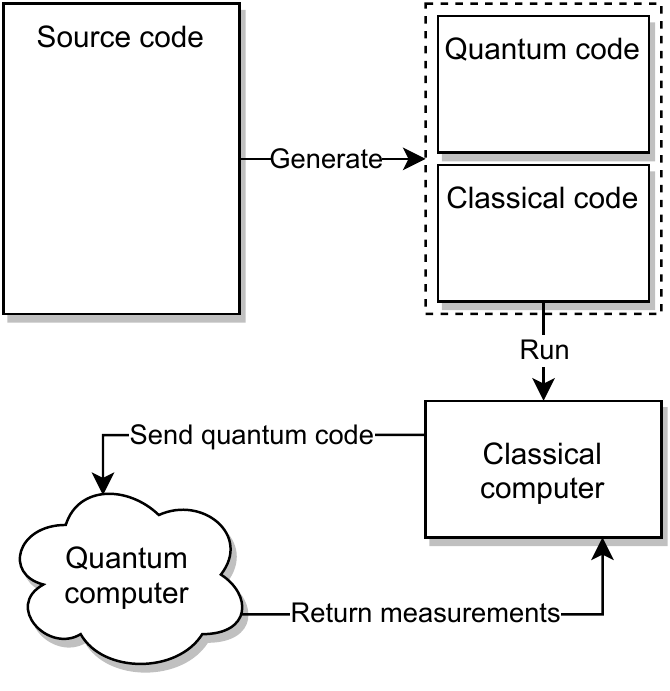}
    \caption{The general workflow of a quantum programming language.}
    \label{fig:lang}
\end{figure}

The characteristics and constraints for a quantum programming language may
diverge from each language. So, for this paper, we will consider the following
ones.

The general workflow of a programming language is summarized in Figure
\ref{fig:lang}.  There, a single source code generates both the classical and
quantum codes.  This generation can be done by a compiler, interpreter (similar
to Python), or a mixed approach. The classical code is then executed by a local
classical computer, and the quantum code is sent to a remote quantum computer.

The quantum computer is a batch computer that takes a quantum code as input and
outputs the measurement results. It is connected to a classical computer
throw a network infrastructure, likely the internet. 

The classical computer coordinates what will be sent to the quantum computer,
setting, \textit{e.g.}, classical parameters of a quantum operation. Because of
performance concerns, the majority of the classical code will be executed
locally.  However, for the quantum code supports the execution of loops
(feedback) and classical controls (feedforward) statements, \textit{e.g.},
\texttt{if-then-else}, the quantum computer needs to be able to execute simple
classical expressions and breaches.

\section{The runtime architecture}
\label{sec:runtime}

\begin{figure*}[ht!]
    \centering
    \includegraphics[width=\linewidth]{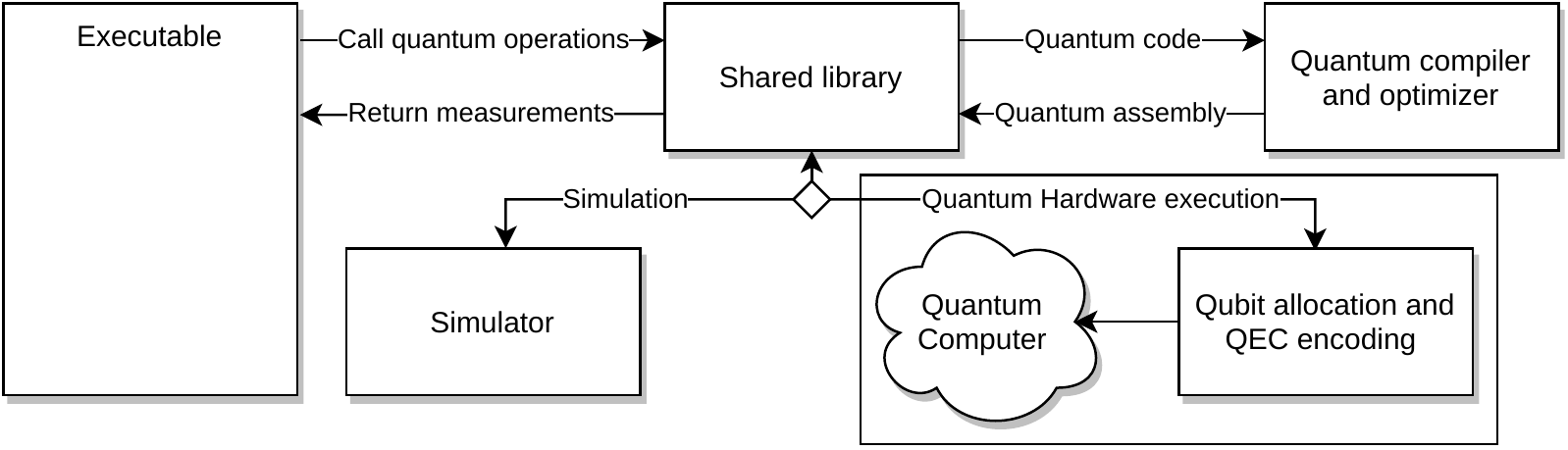}
    \caption{The proposed runtime architecture for dynamic interaction
    between classical and quantum data.}
    \label{fig:runtime}
\end{figure*}

We have designed the runtime architecture, summarized in Figure
\ref{fig:runtime}, following the constraints of Sections \ref{sec:computing}
and \ref{sec:programming}.  The main limitation of the dynamic interaction of
classical and quantum data is the batch execution of the quantum computer.
This problem implies that the classical computer cannot interact with the
quantum computer during its execution, \textit{e.g.}, the classical computer
cannot decide which operation will be executed by the quantum computer based on
some measurement performed on the same execution. This lack of interaction
forbids the execution of some quantum algorithms and protocols, \textit{e.g.},
the quantum teleportation of Figure \ref{fig:tele}. 

To overcome this limitation of interaction, the proposed runtime architecture
relies on quantum computer's ability to execute simple expressions and
breaches, and the generation of the quantum code during the classical runtime.
For further discussions on the implications of this strategy are given in
Section \ref{sec:lib}.

The proposed architecture is composed of several components with the following
workflow. First, an executable makes calls for a shared library that handle the
quantum operations. This executable can be a binary file generated by a
compilation process or an interpreter executing source code like Python. Then
the shared library generates the quantum code during runtime and sends it to a
compiler optimizer that returns a quantum assembly.  After, the quantum
assembly can be sent to a simulator or a quantum computer, where extra steps
are needed to execute. Finally, the results of the quantum assembly execution,
which are the measurement results, are returned to the shared library and then
returned to the executable. The measurement results are classical data and can
be used by the executable as such. 

The following subsections detail every component of the runtime architecture as
well as their interactions and intermediaries. 

\subsection{Shared library}
The shared library works as an intermediary for other components of the runtime
architecture. It handles calls to allocate and free qubits, applies quantum
operations like quantum gates as well as its controlled and inverse versions,
and measurements. 

The calls performed by the executable are used to generate the quantum code,
which is a quantum intermediary representation that is not necessarily
equivalent to a quantum circuit. The quantum code can have higher-level
constructions, such as functions, loops, classical conditional statements, and
arbitrary gates with an unlimited number of control qubits. As the library is
independent of the quantum execution target, the quantum code is independent of
the quantum architecture and has no preset limit of qubits.

To improve performance and coding dynamism, the library shall produce multiple
unrelated quantum codes at the same time. For example, if the parameter of a
quantum application is a random number, and we want to use a real random number
generated by a quantum computer, then we need to initialize a second quantum
code that is independent of the main one, to generate the random number. This
approach reduces the number of qubits, and consequently, gates required to run
the quantum application by splitting it into several quantum executions.

A shared library, in contrast to a static one, is loaded during the execution
and not liked statically into the binary executable. It means that there is no
need to recompile the code on every update, as soon as it maintains the same
interface. Still, a static library can replace the shared library.

\subsection{Quantum compiler and optimizer}

The quantum compiler and optimizer translate the quantum code into a quantum
assembly language and performs optimizations independent of quantum
architecture \cite{Kissinger2019, Nam2018}. The quantum assembly, like the
classical assembly, is a lower-level language, that is closer to the quantum
hardware execution, but it remains independent of the quantum architecture.

For quantum hardware execution on today's NISQ computers, the languages
OpenQASM \cite{Cross2017} and Quill \cite{Smith2016} can be used for,
respectively, IBM and Rigetti quantum computers.  For other executions targets,
such as simulators or resource estimators, other languages \cite{Green2013,
Steiger2018}, and even classical programming languages using quantum
programming libraries \cite{Rosa2020, qiskit, Smith2016, Gidney2018} can be
used. The quantum assembly can be a subset of the quantum code, using the same
quantum programming language.

The quantum assembly must be expressive enough to represent every possible
construction of the quantum code and simple enough to be efficiently executed
by a quantum computer or simulator. It is important to notice that some quantum
programming languages are more expressive than others. For example, the Quil
can represent loops, using label and branch instructions, that is not possible
with the OpenQASM.

\subsection{Quantum simulator}

While we are in the Noise Intermediate-Scale Quantum (NISQ) era, far from large
scale fault-tolerant quantum computers, simulate a quantum computation may be the
best option to test a quantum algorithm or application. However, even with the
arrival of more powerful quantum computers and despite the exponential
simulation time, simulators are yet useful for debugging quantum applications
due to the impossibility of looking into a quantum superposition.

The simulator could execute the quantum code instead of the quantum assembly to
improve performance. Complex operations that need to be decomposed in several
quantum gates to be executed in a quantum computer can be executed in a single
step by a simulator. For example, a controlled-NOT gate with multiple qubits of
control, an oracle for Grover's algorithm \cite{Grover1997}, and the modular
exponentiation for Shor's algorithm \cite{Shor1997} can all be executed by a
single quantum operation in a simulator.

\subsection{Quantum hardware execution}

\begin{figure}[h]
    \centering
    \includegraphics[width=\linewidth]{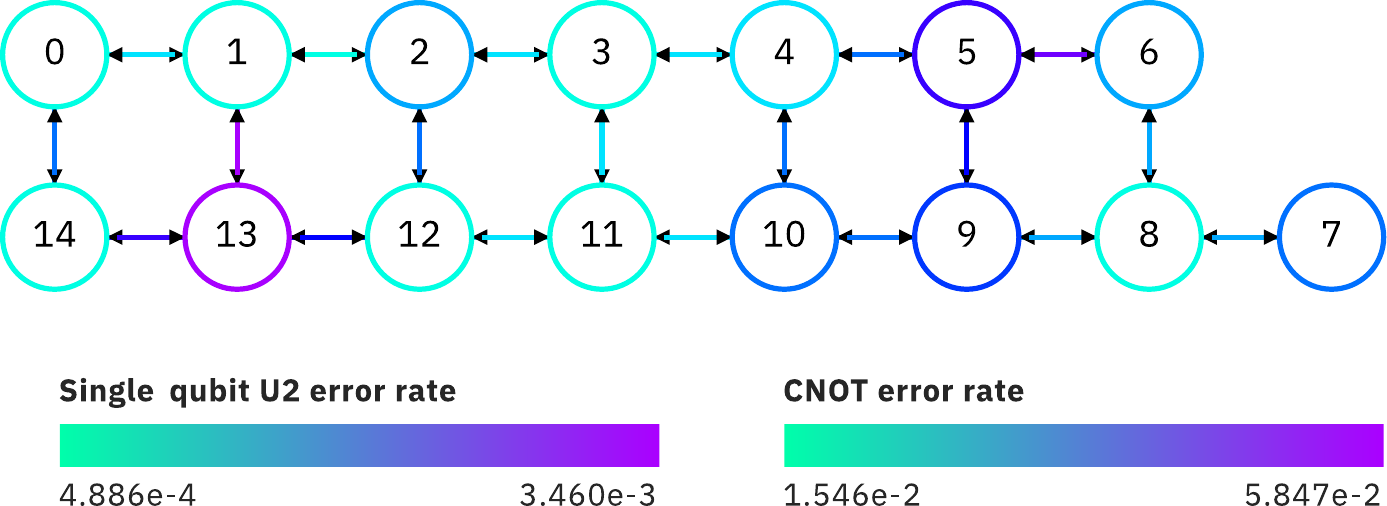}
    \caption{Quantum computer ibmq\_16\_melbourne v2.0.6 calibrated at April
    16, 2020. Screenshot by author from \url{https://quantum-computing.ibm.com}.
    } 
    \label{fig:ibmq}
\end{figure}

The quantum assembly describes the execution of a quantum application in terms
of logical qubits. Those are ideal qubits, free of noise and fully connected,
where no error corrections are needed, and every qubit communicates with each
other, \textit{e.g.}, it is possible to apply a CNOT between any qubit.
However, quantum computers work with physical qubits subject to noise (errors)
and connectivity limitations. An example is the quantum computer IBMQ Melbourne
of 15 qubits. Figure \ref{fig:ibmq} shows its qubit connectivity and
calibration.

The logical qubits of the quantum assembly need to be mapped into physical
qubits to run on a quantum computer. This process is called qubit allocation
\cite{Siraichi2018} or quantum circuit mapping \cite{Itoko2020}. Heuristics for
optimization may vary from each quantum computer vendor since every explores
different qubits layouts and construction technology. 

Despite being too expensive for NISQ computers, quantum error correction is
part fundamental of fault-tolerant quantum computers. So, the quantum assembly
may pass throw a QEC encoder before been send to a quantum computer. As the
qubit allocation, QEC encoding also depends on the quantum architecture and may
be performed by proprietary software provided by the quantum computer vendor.

\section{The shared library features}
\label{sec:lib}

To provide the dynamic interaction between classical and quantum data, the
shared library relies on two main features, the ability to generate the quantum
code at runtime and that a measurement returns a future, in other words, the
promise of a result. With the given restriction of Section \ref{sec:computing}
and \ref{sec:programming},  those two features are essential for the runtime
architecture ability to execute generic quantum applications with the dynamic
interaction between classical and quantum data, and even quantum circuits like
the quantum teleportation of Figure \ref{fig:tele} depends on it.

Due to decoherence, every quantum operation needs to be highly optimized.
Therefore, it is discouraged to construct a sequence of gates that operates
based on an undefined classical input. For example, Shor's algorithm needs to
perform modular exponentiation on a superposition taking a state $\ket{x}\ket0$
into $\ket{x}\ket{a^x \mod n}$, where $a$ and $n$ are classical parameters.
Thus, for the quantum code to be optimized for specific classical parameters,
either they need to be statically defined (compile-time resolvable), or the
quantum code generation needs to be delayed for the runtime where those
parameters are available.

Besides enabling the description of parameterized quantum applications, as
described above, the generation of quantum code at runtime also permits the
description of responsive quantum applications, which can change its execution
based on user input, sensor measurement, or quantum execution. For example, an
application that for a given number returns its prime factors and a quantum
code that uses a measurement result as input.

Some quantum application, like the quantum teleportation of Figure
\ref{fig:tele}, uses the result of a measurement to control which gates are
applied. However, as the quantum computer's decoherence time may be shorter
than the response time between the classical and the quantum computer, a
quantum computer executes in batch. So, such decisions of what needs to be done
based on a quantum measurement demand to be made by the quantum computer.  That
is where measurement results as a future take place.

When a measurement is performed, the shared library returns a promise in the
form of a future to the classical computer. That promise is just fulfilled when
the measurement result is needed by the classical computer and the quantum code
is executed.  This way, classical control statements with quantum operations
are possible in the runtime architecture, since they can be placed in the
quantum code and executed by the quantum computer. A future can hold other
information that is not necessarily a measurement result but is only available
in the quantum computer, \textit{e.g.}, the result of expressions with
measurement results and loop control variables.  In addition to the
\texttt{if-then-else} statement, higher-level construction, such as
\texttt{for} and \texttt{while}, usually available on general propose
programming language, are also possible to be executed by a quantum computer. 

For further programming dynamism, futures should be transparent for the
programmer. For example, the following high-level pseudocode that implements
quantum teleportation
\begin{lstlisting}[numbers=left, xleftmargin=4ex, basicstyle=\linespread{.9}\ttfamily, language=C++,morekeywords={qubit,x,cnot,z,h,measure,print}]
qubit[] bell(aux0, aux1) {
    qubit q0, q1;
    if (aux0) x(q0);
    if (aux1) x(q1);
    h(q0);
    cnot(q0, q1);
    return q0, q1;
}
    
qubit teleport(a) {
    b = bell(0, 0);
    cnot(a, b[0]);
    h(a);
    m0 = measure(a);
    m1 = measure(b[0]);
    if (m1) z(b[1]);
    if (m0) x(b[1]);
    return b[1];
}

int main() {
    qubit a;
    h(a);
    y = teleport(a);
    print(measure(y));
}
\end{lstlisting}
generates the fowling quantum code:
\begin{lstlisting}[numbers=left, xleftmargin=4ex, basicstyle=\linespread{.9}\ttfamily, language=C++,morekeywords={qubit,x,cnot,z,h,measure,print}]
qubit a;
h(a);
// teleport begin
// bell begin
qubit q0, q1;
h(q0);
cnot(q0, q1);
// bell end
cnot(a, q0);
h(a);
m0 = measure(a);
m1 = measure(q0);
if (m1) z(q1);
if (m0) x(q1);
// teleport end
return measure(q1)
\end{lstlisting}

Note that the \texttt{if} statements of the function \texttt{bell} are not
present in the quantum code since the values \texttt{aux0} and \texttt{aux1}
are known by the classical computer.  However, the \texttt{if} statements of
the function \texttt{teleport} operate with measures that are unknown by the
classic computer, so they are placed in the quantum code. The quantum code is
executed just when the measurement of the qubit \texttt{y} is needed for the
print function in the main function. 

Every future is a node of an abstract syntax tree (AST) that is generated at
runtime. An AST is a data structure represented by a tree that stores the
syntax of a program, in our case, the syntax of the quantum code. An operation
between a future and some classical data also generates a new node (future),
and the results are just placed into the futures (the promise is fulfilled)
after the quantum code execution.  When the classical computer requests the
result of a future, the AST is traversed, beginning at the node where the
result was requested, generating the quantum code. Every qubit also has its own
AST holding the gates applied on it,  which the future returned from the qubit
measurement is connected as a parent node.

The generation of quantum code at runtime and the ability to return a future on
a measurement allows that quantum and classical information interact regardless
of where they are stored. This interaction can be in terms of (i) classical
data controlling the application of quantum operations, (ii) quantum
measurement results used as input of a classical function, and the loop of
those two, all transparent to the programmer.

\section{Conclusion}
\label{sec:conclusion}

We presented a runtime architecture for the constructions of quantum
programming languages that enable dynamic interaction between classical and
quantum data. Considering the restriction that a quantum computer processes in
batch, making impossible the interaction between quantum and classical computer
during the quantum execution, we propose the use of futures and the generation
of quantum code at runtime to work around this limitation. 

A future holds classical information that is generated in the quantum computer.
It is used as a promise for a measurement result that is fulfilled after the
quantum execution. Futures operates in both the classical and the quantum
computer transparent for the programmer. With this strategy, classical
constructions like \texttt{if-then-else}, \texttt{for}, and \texttt{while} are
also possible to be executed in a quantum computer. 

It is important to notice that full support for the classical constructions
depends on the quantum target and quantum assembly language used.  However,
taking it into account, the loop of classical information controlling the
execution of quantum operations and the result of quantum executions operating
with classical data is possible on any quantum computer, including NISQ
computes. The proposed runtime architecture also addresses both simulation and
quantum computer execution considering its uniqueness.

Embed a quantum programming language on a general-propose programming language
is a widely used strategy, which is supported by the proposed runtime
architecture. However, it is important to notice that semantic checks during
compilation time can be difficult or impossible to implement depending on the
host language and that some constructions can seem out of place with the host
language.

Our runtime architecture presents a general framework for a quantum programming
language and its programming environment. However, we do not properly address
the problem of debugging quantum programs, and there is a lack of precision in
the description of the quantum hardware execution.  As future works, we intend
to use this runtime architecture in the development of new quantum programming
language, approaching the dynamic interaction between classical and quantum
data. 

\section{Acknowledgements}
This study was financed in part by the ``Coordenação de Aperfeiçoamento de
Pessoal de Nível Superior'' - Brazil (CAPES) -  Finance Code 001.

\bibliographystyle{elsarticle-num-names}
\biboptions{sort&compress}
\bibliography{main}

\end{document}